# Multiple Scattering Effects in Noctilucent Clouds:
# Numerical Estimation and Application to Altitude and Particle Size Measurements

Oleg S. Ugolnikov

Space Research Institute, Russian Academy of Sciences, Moscow, Russia
E-mail: ougolnikov@gmail.com

**Abstract**

The simple 3-D radiative transfer model in the atmosphere of the Earth is built for numerical comparison of direct solar radiation and limb scattering background at the definite layer during the deep twilight period at the middle and upper atmosphere. This allows finding the contribution of multiple scattering in the field of high-altitude clouds such as polar stratospheric and noctilucent clouds. This is the factor that can influence the results of altitude and microphysical study of the particles based on the approximation of single scattering. The possible errors caused by this effect are estimated together with the contribution of multiple scattering for different altitudes of the clouds, wavelengths and solar zenith angles. The results are interpreted geometrically and optically, color effects observed for noctilucent clouds are described.

## 1. Introduction

Since the discovery in late 19[th] century (Leslie, 1885), noctilucent clouds (NLC), the highest above the Earth, were considered as the indicator of the climate changes in the entire atmosphere. Possible relation with radiative cooling on $CO_2$ molecules (Roble, Dickinson, 1989) and atmosphere shrinking effect (Berger, Lübken, 2011; Lübken et al., 2018) makes the long-time altitude measurements important, while the increase of mesospheric $H_2O$ can be displayed as the positive trend of mean particle size and NLC brightness. The particle size and the altitude can be measured by different techniques – rocket, active lidar sounding, remote sensing from the ground and satellites.

Optical passive sensing of NLC is often based on the measurements of solar radiation scattered on NLC particles. Since the particle size is less than optical wavelengths, scattering is well described by Mie theory (Kokhanovsky, 2005), however, non-spherical effects are also seen (Baumgarten et al, 2002). Angular dependencies of scattered radiation measured in different spectral bands can be used for the particle size determination as averaged on the cloud ensemble (Ugolnikov et al., 2017) or for different fragments separately (Ugolnikov, 2024). Color or spectral measurements of NLC can also be used for single-site altitude determination of the clouds as averaged (Ugolnikov, 2023a) or across the wave pattern of NLC (Ugolnikov, 2023b). The technique is based on color changes of the clouds as they immersed into the shadow of stratospheric ozone and then dense tropospheric layers, those make NLC bluer and redder, respectively. This effect is often seen by naked eye or fixed by photographs, see Figure 1.

NLC colorimetric altitudes are in good agreement with the triangulation data on the same ensembles of clouds (Ugolnikov et al., 2025). Both techniques refer to the brightest fraction of clouds consisting of largest particles. Colorimetric and triangulation altitudes can be directly compared with historical triangulation measurements for the study of long-term trends possibly originated by the atmosphere shrinking effect. Theoretical estimation of the trend value (Lübken et al., 2018) is – (1-2) km/century. The mean optical altitude of bright NLC in 2020-2024 is about 81 km (Ugolnikov et al., 2025). This is lower than historical triangulation altitudes. Lidar backscatter altitudes referring to the smaller particles are about 2 km higher (Gerding et al., 2021). No difference is seen if we compare the modern lidar and the historical triangulation altitudes (von Zahn, 2003).

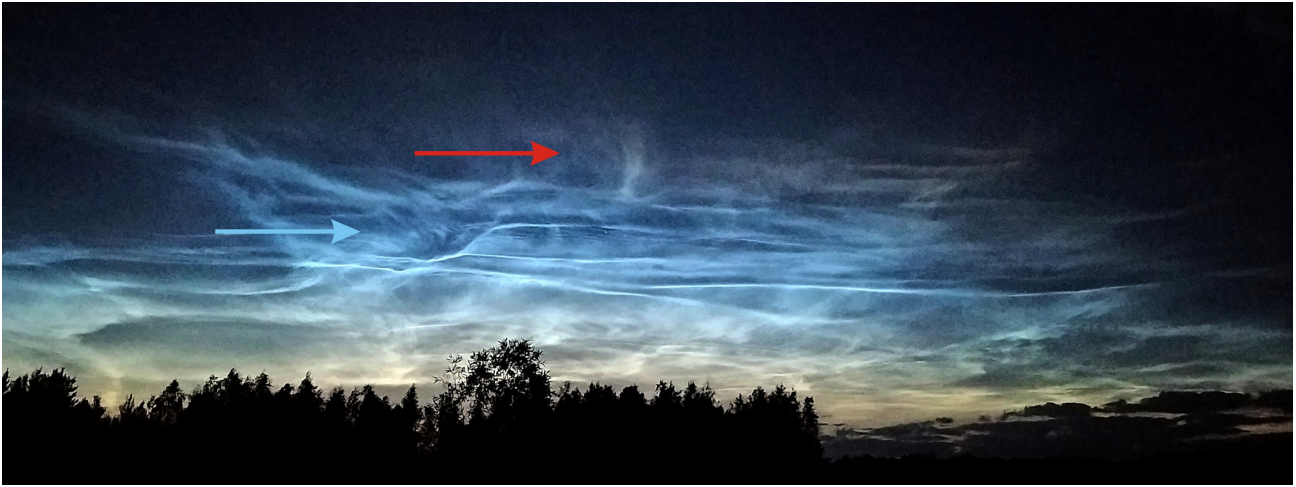

*Figure 1. Color changes of noctilucent clouds near the shadow of the Earth. The red arrow shows the regions illuminated through the troposphere, the blue arrow reflects the shadow of ozone.*

Altitude and particle size color measurements are based on the model of single scattering of solar emission by NLC particles. Strong polarization of NLC radiance at 90° from the Sun (Ugolnikov et al., 2016; Ugolnikov and Maslov, 2019) and clear visibility of the ozone/troposphere/limb shadow effects (Figure 1) point to the domination of single scattering. However, models (Lange et al., 2022) show that multiple scattering can be significant, especially in the blue spectral range, this can influence the results of NLC analysis described above.

It is known that the contribution of multiple scattering in the sky background during the twilight stage when NLC are observed sufficiently increases and becomes dominant (Ugolnikov, 1999; Ugolnikov and Maslov, 2002, 2013), but it occurs mainly in the lower atmosphere (Patat et al., 2006). Multiple scattering is also the problem for limb satellite analysis of atmospheric species and aerosol (Oikarinen et al., 1999). Radiative transfer models were developed for analysis of satellite mission data. Pseudo-3D model LIMBTRAN (Griffoen and Oikarinen, 2000) was presented prior to the launch of Odin satellite with Optical Spectrograph and Infrared Imager System (OSIRIS, Llewellyn et al., 2004); model GOMETRAN (Rozanov et al., 1997) was designed for Global Ozone Monitoring Experiment (GOME, Burrows et al., 1999) onboard ERS-2 satellite. It was updated to SCIATRAN model (Rozanov et al., 2017) used for NLC analysis by Lange et al. (2022). The vector model MCC++ (Postylyakov, 2004) describes the behavior of intensity and polarization of the sky background during the light period of twilight (Ugolnikov et al., 2004).

Noctilucent clouds are visible in the sky during the deep stage of twilight, at solar zenith angles starting from 95°. Cloud is being immersed into the shadow and fades at local solar zenith angle ζ (visible from the cloud) less than 99°. When the Sun is deeper below the horizon, we can see NLC only in the dusk/dawn segment, where local solar zenith angle is less than solar zenith angle in the observation site.

Modelling of radiative transfer during the dark stage of twilight with unlimited order of scattering is the complicated problem of numerical analysis. Errors can increase if we fix the sky field component (NLC as the example) by subtraction of computed fields with and without this component. The problem can be simplified if the model is designed for the estimation of the fixed characteristics not building the complete picture of the radiation field. In this paper we find the ratio of single and multiple scattering in the NLC field for the fixed wavelength by the single model run. Since NLC are optically thin, these components can be interpreted by scattering of direct solar radiation and limb background by NLC particles. To fix this, we have to compare the total intensity of the Sun and the limb radiance depending on the solar zenith angle and to estimate the effects caused by difference of solar and limb positions and the difference of corresponding angles of scattering. This allows the estimation of the multiple scattering effects on the results of NLC altitude and particle size retrieval.

## 2. Radiative transfer model

### 2.1. Basic principles

Monte-Carlo simulation of radiative transfer (Marchuk et al., 1980) is an effective tool, especially if the general property such as the total limb radiance is being studied. Forward and backward photon propagation can be simulated. The backward approach (Collins et al., 1972) is often used in the case of the fixed detector with the narrow field of view (SIRO algorithm, Oikarinen et al., 1999, for example). In our case, we calculate the integral radiance from the limb simultaneously from different locations in NLC layer for the different local solar zenith angles. This defines the choice of the forward Monte-Carlo simulation.

We don't express the noctilucent clouds as the uniform aerosol layer with fixed properties, those are actually various and previously unknown. However, we can solve this problem for different altitudes of clouds, and the results can be used for other cloud layers such as polar stratospheric clouds (20-25 km) and recently observed winter noctilucent clouds (70 km, Ugolnikov, 2025).

Running the model, we find the values of direct solar emission and total limb radiance for the atmospheric point with the altitude $h$ and local solar zenith angle observed from the cloud, $\zeta$, for different wavelengths $\lambda$. The Sun/limb intensity ratio will be close to the ratio of single and multiple scattering by the NLC element in this atmospheric point with possible corrections described below.

To find the straight solar and limb intensities as the functions of $\zeta$ and $h$ for the fixed wavelength $\lambda$, we define the Cartesian geocentric coordinate system with the axis $x$ directed to the anti-solar point (see Figure 2). Following (Ugolnikov, 2023a), we take the Earth's radius $R_E$ equal to the curvature radius of Earth's terminator for the summer deep twilight conditions at the altitudes 55°-60°N (6385 km). Initially we consider the Sun as the point source, its angular size and the effect of darkening to the edge are taken into account at the following stage. Solar radiation enters the Earth's atmosphere along the lines $y$ = const, $z$ = const. Due to the spherical symmetry, we can calculate the direct solar photon flux $P_S(h, \zeta)$ and limb background photon flux $P_M(h, \zeta)$ integrating over the circles with the fixed $h$ and $\zeta$:

$$h = \sqrt{x^2 + y^2 + z^2} - R_E; \quad \zeta = \frac{\pi}{2} + \arctan \frac{x}{\sqrt{y^2 + z^2}}.$$

(1)

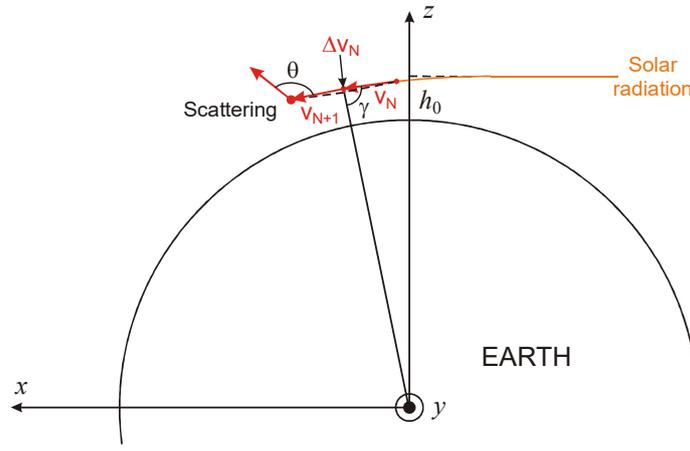

*Figure 2. Coordinate system and scheme of the steps of photon propagation analysis.*

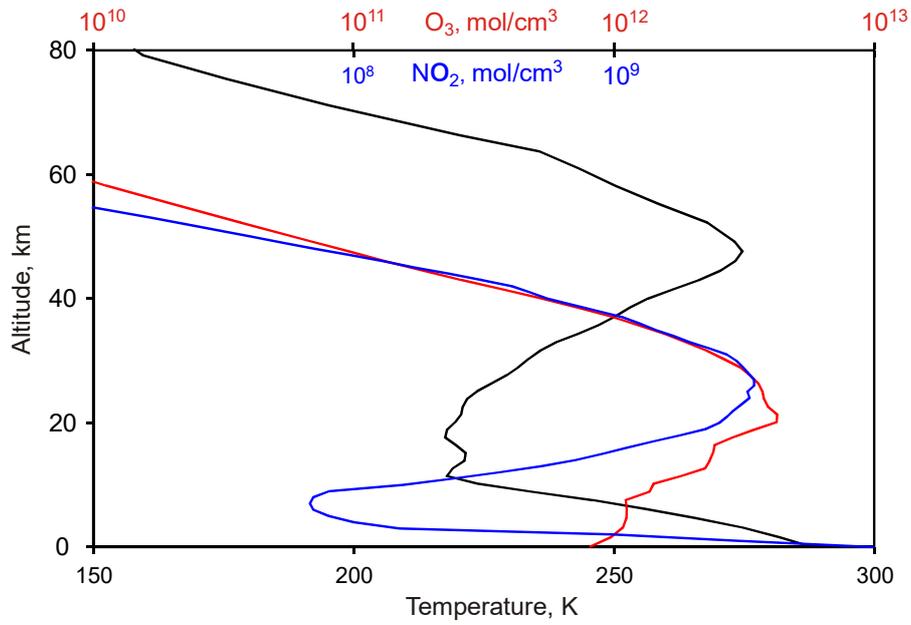

*Figure 3. The profiles of temperature, ozone, and nitric dioxide ($\zeta = 90°$) used in the model.*

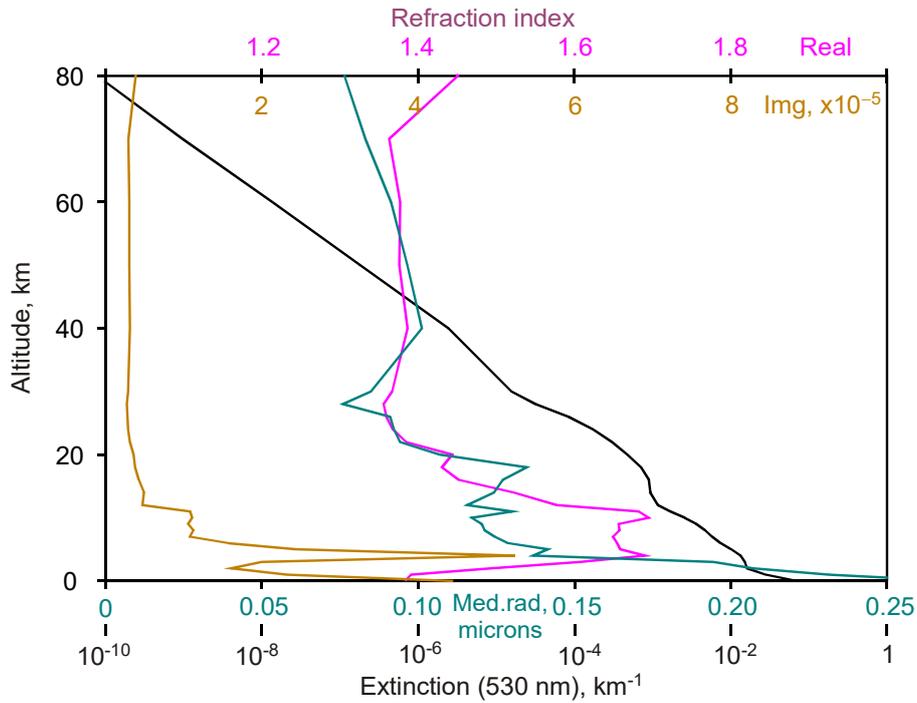

*Figure 4. Characteristics of atmospheric aerosol used in the model.*

Owing to the symmetry, we can fix $y = 0$ for all incoming solar photons; $z$-coordinate is variable. It will not change the sums $P_S$ and $P_M$. However, $y$-coordinate of the photon can be changed along the trajectory as the result of scattering or reflection from the Earth's surface. Monte-Carlo simulations of radiative transfer with account of refraction, gas and aerosol absorption, and Rayleigh and aerosol scattering are used for retrieval of the functions $P_S(h, \zeta)$ and $P_M(h, \zeta)$.

## 2.2. Model data

The model is designed for the estimation of multiple scattering in the observed field of noctilucent clouds, which can affect the results on particle size and altitude. Ground-based observations are usually held in wide spectral bands in the visible spectral range. The basic factors those must be taken into account (Ugolnikov, 2023a), are following: Rayleigh scattering; aerosol scattering and absorption; wide absorption bands of $O_3$ and $NO_2$ in the optical range. Temperature profile defining the altitude dependence of Rayleigh scattering and the profile of ozone are taken from EOS Aura/MLS satellite data (NASA GES DISC, 2025) as averaged for the dates of NLC observations in 2020-2024 (Ugolnikov et al., 2025), profiles are plotted in Figure 3. The vertical profile of $NO_2$ is taken as function of local solar zenith angle $\zeta$ according to (Gruzdev, Elokhov, 2021); evening and morning profiles are averaged. The terminator profile ($\zeta = 90°$) of $NO_2$ is also shown in Figure 3.

The most complicated is the model of aerosol in the troposphere and stratosphere. Figure 4 shows the profiles of the real and imaginary parts of the particle refraction index taken from the model (Zuev, Krekov, 1986). The particle size distribution is assumed to be lognormal with constant width 1.6, the profile of the median radius is also plotted in the figure. Total particle number is defined by extinction profile at the wavelength 530 nm. It is corrected using the recent OMPS limb data (NASA GES DISC, 2025) averaged over the same dates of NLC observations during the last 5 years, it is also shown. Extinction and scattering coefficients, scattering angle dependence for the given altitude and wavelength are calculated by Mie theory.

## 2.3. Numerical procedure

We consider the trajectory of the solar photon entering the atmosphere close to the Earth's terminator. It is defined by initial position $\mathbf{r_0} = (x_0, 0, z_0)$ and normal vector of propagation $\mathbf{v_0} = (1, 0, 0)$. The value of $x_0$ is equal to $-1200$ km, that is enough for the trajectory to start outside the atmosphere, $z_0$ is equal to $R_E + h_0$, where $h_0$ randomly varies from $-50$ to $+101$ km. The lower boundary value is chosen by the analysis of the final model results: photons with $h_0 < -50$ km do not significantly contribute to the limb radiance. The photon is characterized by its elemental weight that is close to unity, but here we can take into account the tiny increase of photon number for larger $h_0$ due to geometrical factor, and define the initial photon weight as

$$J_0 = 1 + \frac{h_0}{R_E}$$

(2)

The step of the numerical procedure is $s = 1$ km. The coordinate of the photon changes during the step:

$$\mathbf{r_{N+1}} = \mathbf{r_N} + s \cdot \mathbf{v_N}$$

(3)

The photon experiences refraction at this step. Photon trajectory is deflected along the Earth's radius by an angle

$$\Delta\alpha = s \sin\gamma \frac{d\mu(h)}{dh}$$

(4)

Here $\gamma$ is an angle between the trajectory and the direction to the center of the Earth, $d\mu/dh$ is the derivative of the refraction index of the gas medium on the altitude. This derivative is negative, so the refraction angle, the photon is deflected to the center of the Earth. We take the small vector

$$\Delta\mathbf{v_N} = s \cdot \frac{\mathbf{r_N}}{r_N} \cdot \frac{d\mu}{dh}$$

(5)

In the case of small angle $\Delta\alpha$ the angle between vectors $\mathbf{v_N}$ and $\mathbf{v_N} + \Delta\mathbf{v_N}$ is equal to $\Delta\alpha$ for any value of $\gamma$ (see Figure 2). This allows finding the new photon direction with account of refraction by the least-computation procedure:

$$\mathbf{v_{N+1}} = \frac{\mathbf{v_N} + \Delta\mathbf{v_N}}{|\mathbf{v_N} + \Delta\mathbf{v_N}|}$$

(6)

The new direction vector is normalized by division by its module. Along with the refraction, scattering and absorption take place. As it is usually done in Monte-Carlo procedures, we consider them separately: the absorption decreases the photon weight not changing its direction, while scattering change the photon direction with the constant weight. Absorption is not being considered as a random process. It is the sum of aerosol, ozone, and nitrogen dioxide absorption:

$$\ln J_{N+1} = \ln J_N - s\left(A(h,\lambda) + \sigma_{O3}(\lambda)\cdot n_{O3}(h) + \sigma_{NO2}(\lambda)\cdot n_{NO2}(h,\zeta)\right)$$

(7)

Here $A$ is the absorption coefficient of aerosol calculated from the model shown in Figure 4, $\sigma_{O3,NO2}$ and $n_{O3,NO2}$ – cross sections and concentrations of ozone and nitrogen dioxide. Rayleigh and aerosol scattering can occur, they are also considered separately. Probabilities of Rayleigh and aerosol scattering are:

$$p_{R,A} = s\cdot C_{R,A}(h,\lambda)$$

(8)

Here $C_{R,A}$ means Rayleigh and aerosol scattering coefficients. Scattering is modeled as random process taking into account the Rayleigh and aerosol scattering angle dependencies. The last is calculated by Mie theory basing on the particle size distribution and refraction index for the fixed altitude. A similar procedure is executed if the photon reaches the surface of the Earth. In this case its weight is multiplied on the albedo that is taken as constant value (0.367), after that the photon is being randomly redirected to the sky hemisphere.

As the step is modeled, the values of the photon altitude $h$ and local solar zenith angle $\zeta$ are found. If the photon is on the twilight/night side of the Earth ($x>0$, $\zeta>90°$), and the altitude $h$ is inside the interval from 10 to 100 km, its weight is added to the sums:

$$P_S(h,\zeta) \rightarrow P_S(h,\zeta) + J_{N+1},$$

if no scattering or reflection from the Earth's surface had occurred with this photon;

$$P_M(h,\zeta) \rightarrow P_M(h,\zeta) + J_{N+1},$$

if at least one event of scattering or reflection had occurred. (9)

It is worthy of note that the number of photons fixed in the cell ($\Delta h = 1$ km, $\Delta\zeta = 0.1°$) is proportional to the photon density and the volume of the cell, not depending on the shape of the cell and direction of photon propagation. The procedure cannot be halted as the result of photon absorption or reaching the Earth's surface, its weight is reduced in these cases. The process is finished if photon escapes the atmosphere, its altitude becomes higher than 120 km and continues increasing. We also see that the direct solar and limb radiance are calculated simultaneously for different altitudes, and there is no need to restart it if another atmospheric layer is studied. This helps to reach good output accuracy reducing the calculation time.

Results of this Monte-Carlo procedure are the distributions $P_S(h, \zeta)$ and $P_M(h, \zeta)$ for the case of the point source. We have to take into account the angular size of the Sun with the radius $\rho_0$:

$$I_{S,M}(h,\zeta) = \int_{-\rho_0}^{\rho_0} S(\rho)P_{S,M}(h,\zeta+\rho)d\rho \cdot \frac{1}{\sin\zeta\cdot(1+h/R_E)}; \quad \int_{-\rho_0}^{\rho_0} S(\rho)d\rho = 1$$

(10)

Here the function S(ρ) is the vertical section of solar disk brightness. It slightly depends on the wavelength owing to different laws of darkening of the solar disk to the edge. The factor $(\sin\zeta \cdot (1+h/R_E))$ is close to unity, it reflects the difference of the volumes of the circular cells restricted by altitude ($h$, $h+\Delta h$) and solar zenith angle ($\zeta$, $\zeta+\Delta\zeta$). Functions of $I_S(h, \zeta, \lambda)$ and $I_M(h, \zeta, \lambda)$ are used to analyze the contribution of multiple scattering in the field of high-altitude clouds for different heights, solar zenith angles and wavelengths. Effects of multiple scattering on the altitude and particles sizes based on single scattering models are also reviewed.

## 3. Results

Figure 5 shows the dependencies of $I_S(\zeta)$ and $I_M(\zeta)$ for the altitudes 20 km (typical for polar stratospheric clouds), 45 km, 70 km (winter noctilucent clouds), and 82 km (classical noctilucent clouds) for the wavelength 530 nm. The same dependencies for the wavelengths 440 and 620 nm are added for 20 and 82 km. All brightness values are normalized on straight unabsorbed solar radiation, so $I_S \rightarrow 1$ at sunrise/sunset in the upper atmospheric layers. The dependencies are easy to be interpreted. Direct solar radiation remains constant until the Sun starts immersing into the shadow of dense atmospheric layers, while limb radiance gradually decreases as the Sun approaches the limb. The dependency of limb radiance $I_M(\zeta)$ is close to exponential, especially for high altitudes corresponding to noctilucent clouds. It is worthy of note that the gradient of the limb radiance logarithm on the solar zenith angle for NLC is practically independent on the wavelength, being close to

$$d \log I_M / d\zeta \approx -0.2/° \tag{11}$$

The main result of the calculations is very small ratio of multiple scattering in total radiance $I_M/I_T = I_M/(I_S+I_M)$ for noctilucent clouds when they are above the shadow of dense atmospheric layers, as we see in Figure 6. During the twilight period with solar zenith angles from 95° to 98°, when NLC are easily observed in the sky, the limb radiance is weaker than 1% of the Sun. Low contribution of multiple scattering is confirmed by clear shadow boundary visible in bright NLC fields and blue and red color lines near the shadow described above.

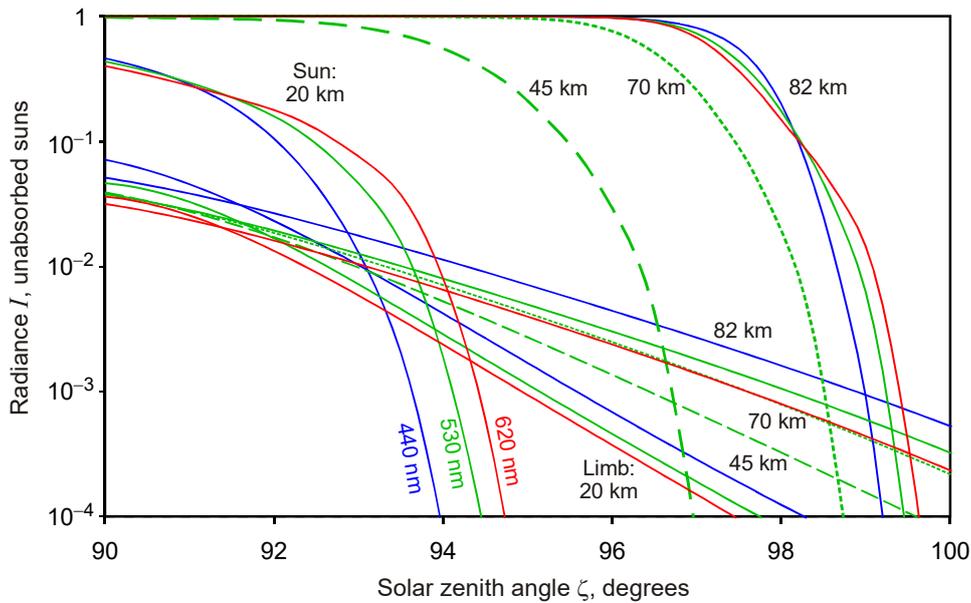

*Figure 5. Direct solar and total limb radiances depending on the solar zenith angle at the altitudes 20, 45, 70, and 82 km for the wavelengths 440, 530, and 620 nm. At 45 and 70 km, only the dependencies for 530 nm are shown (dashed and dotted lines, respectively).*

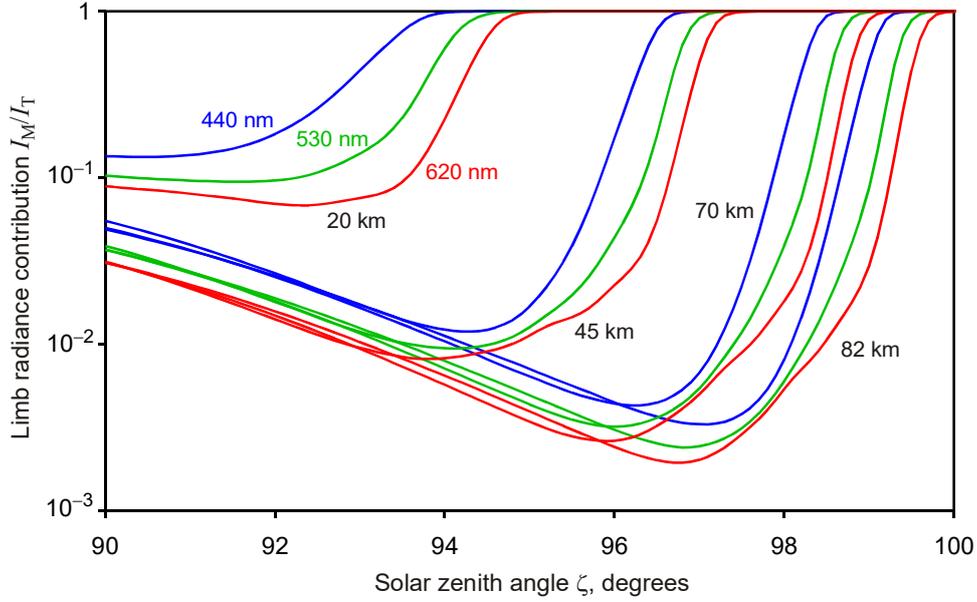

*Figure 6. Contribution of the limb in the total radiance for the same altitudes and wavelengths as in Figure 5.*

Multiple scattering becomes more significant in the lower layers of the atmosphere. For polar stratospheric clouds it is not less than 10% in the green-yellow spectral range during the period of best visibility (solar zenith angle 91°-92°), increasing to 20% in the blue spectral range. As the Sun is depressing, contribution of multiple scattering becomes higher. It is known that polar stratospheric clouds can remain visible during the dark twilight period being immersed in the shadow of the Earth, the effect is never seen for noctilucent clouds.

An evolution of illumination conditions of the high-altitude clouds during the twilight is clearly seen if we draw the dependencies of color ratios $I_T(\lambda_2) / I_T(\lambda_1)$ on the solar zenith angle. They are shown in Figure 7 for the wavelength pair (530 nm to 440 nm), the ratios for single and multiple scattering are also drawn. Here the intensities are normalized on unabsorbed Sun, the same in all wavelengths ("white Sun model"), and the color ratio is close to unity during the light twilight. Real cloud color with account of solar spectrum, particle scattering properties, and local atmospheric transparency differs by constant factor, but the dependencies remain the same. This will be described below.

Color evolution of noctilucent clouds was described by Ugolnikov (2023a), these effects are also seen here. Color ratio decreases (the color turns bluer) when the clouds are immersed in the shadow of the ozone layer, for NLC this corresponds to the solar zenith angle 97°-98° and seen in Figure 7. After that the color turns redder owing to the depression of the Sun beyond the dense tropospheric layers. These color dependencies are in good agreement with ground-based photometry and were used to estimate the altitude of NLC (Ugolnikov, 2023a). Polar stratospheric clouds is immersed into the shadow at the solar zenith angle 93°-94°, but then the color turns back bluer owing to the domination of multiple scattering.

As we can see in color ratio dependencies, limb radiance has a little excess in the blue part of spectrum, basically owing to Rayleigh scattering along the tangent direction of light propagation. For NLC, multiple scattering becomes principal just at the local solar zenith angle 99°, when NLC are in fact immersed into the shadow of the Earth and are not observed and analyzed. At solar zenith angles 95°-98.5°, where observations of NLC are made, multiple scattering is a small admixture. In the following chapters, we analyze its effects on altitude and particle size measurement results based on single scattering approximation.

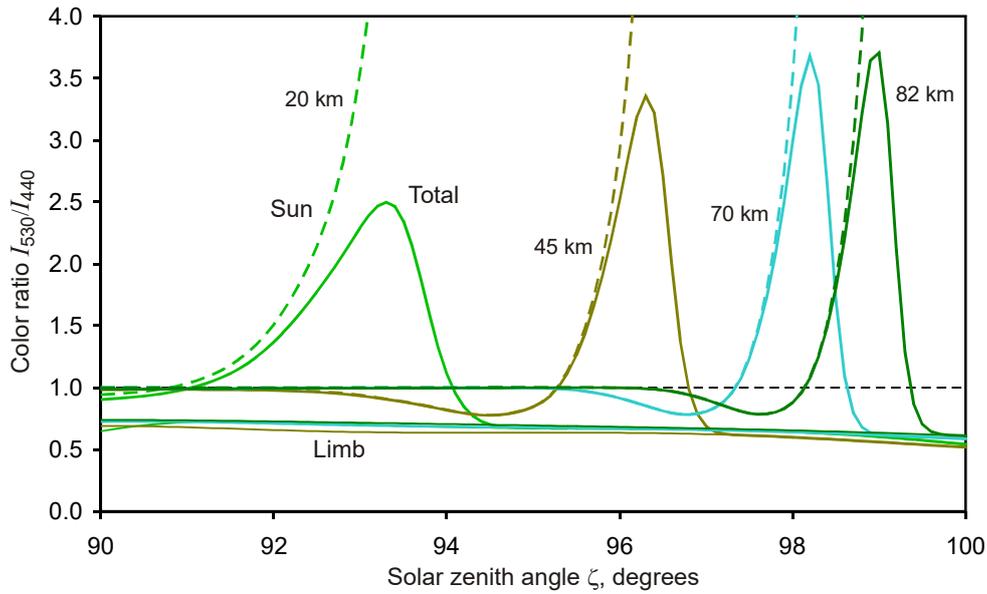

*Figure 7. The color ratio (530 to 440 nm) of cloud illumination at the altitudes 20, 45, 70, and 82 km, "white Sun" model.*

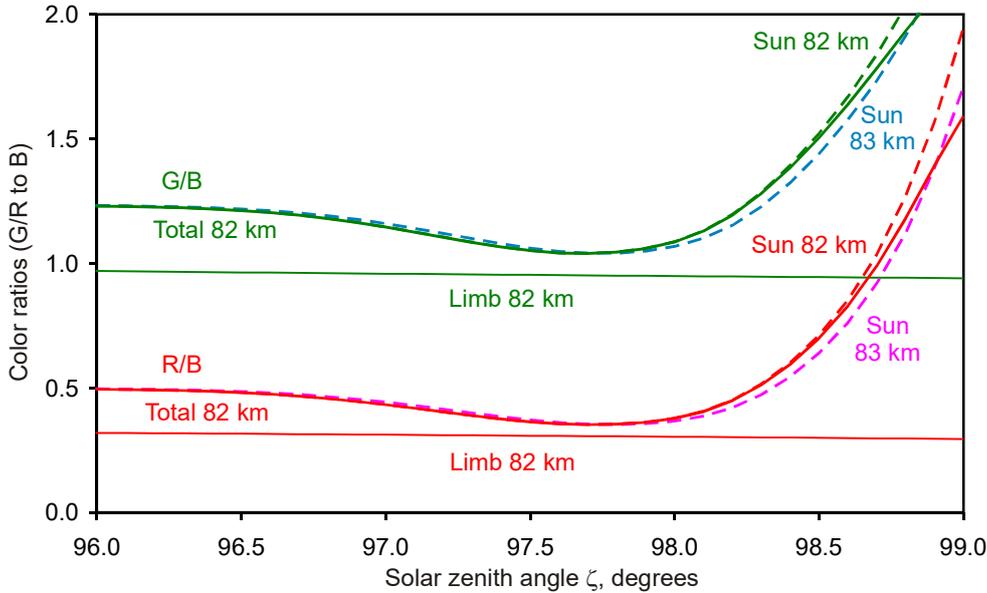

*Figure 8. The RGB color ratios of solar emission, limb radiance, and total background at 82 km during the period of NLC visibility with account of solar spectrum, NLC light scattering and atmospheric transparencies. The color ratio of solar emission is also shown for 83 km.*

## 4. Effects in altitude estimation

Change of NLC color near the Earth's shadow border caused by extinction of tangent solar emission in dense atmospheric layers can be the basis of altitude estimation, if we know the altitude profiles of absorption and scattering as the functions of wavelength. This single-camera technique was used by Ugolnikov (2023ab) to find the mean altitude of NLC, the same method allowed finding the altitude of winter noctilucent clouds observed for the first time (Ugolnikov, 2025). The single scattering model was used in all cases. Comparison with triangulation data for the same NLC ensembles in 2020-2025 (Ugolnikov, 2024; Ugolnikov et al., 2025) had shown a good agreement, however, photometric altitude slightly exceeds the triangulation altitude, the mean difference is about 0.5 km. Photometric altitude of winter noctilucent clouds in December 2024, 70 km, was also 1.5 km above the deep temperature minimum by MLS and SABER data, however, the error of altitude estimation was of the same value in that case. This was interpreted as a possible effect of multiple scattering: owing to excess in the blue part of spectrum, it slows the reddening of NLC near the shadow border.

Figure 8 presents a large-scale view of color ratios of noctilucent clouds during the visibility period. It shows the modeled color ratios (G/B) and (R/B) for RGB color device used by measurements (Ugolnikov, 2023ab). Here the solar spectrum, NLC scattering properties and local atmospheric transparency were taken into account as in referred papers, these colors are confirmed by the measurements. We plot the model colors for 82 km together with single scattering colors for 82 and 83 km. We see the tiny "bluing effect" caused by multiple scattering, its value significantly depends on solar zenith angle. Fitting the total color by single scattering colors in the observational interval of solar zenith angles from 95.0° to 98.6° (Ugolnikov, 2023a), we find the least squares altitude: 82.15 km. This allows fixing the systematical shift of these altitude measurements, it is not more than about 0.2 km in visible spectral range. The shift is even less if near-umbra regions of NLC with $\zeta \sim 98.5°$ are not analyzed.

## 5. Effects in particle size estimation

Statistical analysis of the directions of the photons scattered in the lower atmospheric layers relatively the sunward direction allows building the image of the limb as observed from the noctilucent cloud at the altitude 82 km. It is shown in Figure 9 for different solar zenith angles; blue, green, and red colors in the figure correspond to the wavelengths 440, 530, and 620 nm, respectively. We see that the limb radiance forms the narrow crescent surrounding the Sun position, it turns narrower as the Sun sets behind the limb.

If we take the definite position of observer on the Earth, the single scattering angle $\theta_S$ will be equal to the angle between the directions to the cloud and to the Sun with the small correction caused by the atmospheric refraction. If the photon comes from the limb, scattering angle will differ from the case of direct solar radiation. However, as we see in Figure 9, the Sun is close to the middle of brightest part of the limb, effective scattering angle $\theta_M$ is expected to be close to the single scattering angle.

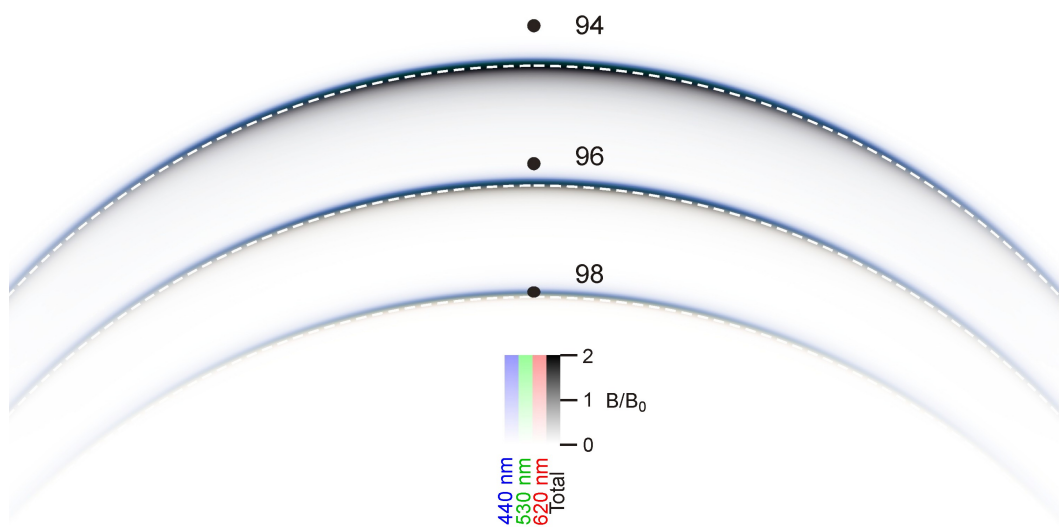

*Figure 9. Limb profiles visible from the altitude 82 km for different solar zenith angles (negative). Positions of the Sun and limb (white dashed line) are shown with account of refraction.*

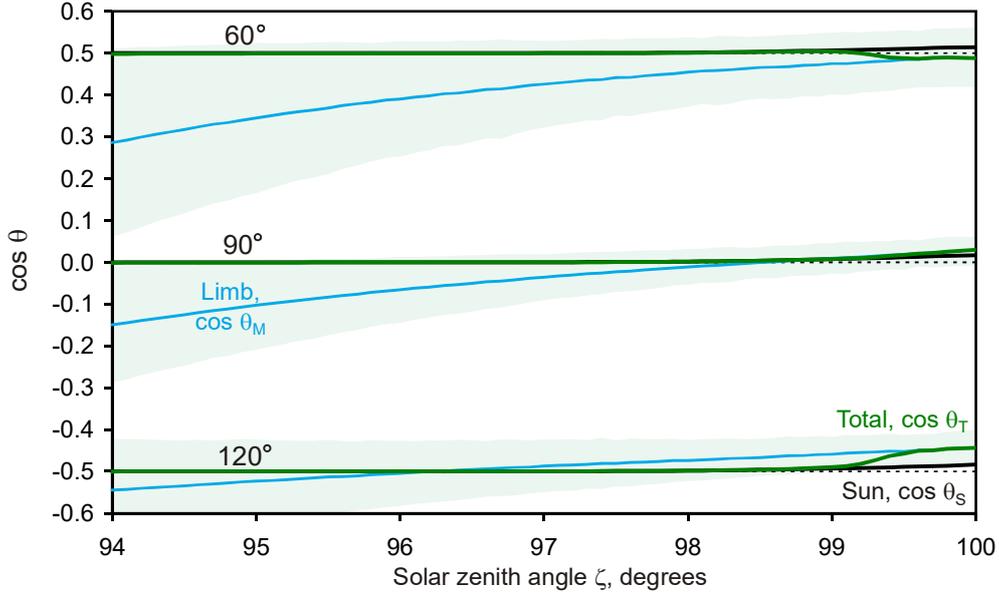

*Figure 10. Effective cosines of scattering angle for solar vertical points 60°, 90°, and 120° from the Sun for direct solar emission, limb radiance, and the total background, 530 nm.*

The method of mean particle radius estimation was suggested by Ugolnikov et al (2017) and developed by Ugolnikov (2023ab; 2024). It is based on the measurements of the color ratio of NLC field at different scattering angles for the same solar zenith angle visible from the cloud, not from the observer. For particle sizes less than 100 nm the color ratios change linearly with $\cos \theta_S$, the coefficient is proportional to the square of the mean particle radius $a^2$.

If scattering of limb background occurs, the difference between the effective scattering angle $\theta_T$ and the single scattering angle $\theta_S$ appears. To estimate this effect numerically, we take three directions in solar vertical in 60°, 90°, and 120° from the Sun and calculate the mean value of $\cos \theta$ for direct solar radiation, limb background, and the total radiance. The results are shown in Figure 10 for wavelengths 440, 530, and 620 nm. As spoken above, the mean $\cos \theta_S$ is equal to 0.5, 0, and –0.5 for these three directions, respectively, with a tiny correction by refraction. Mean $\cos \theta_M$ value of multiple scattering is below $\cos \theta_S$ for the early stage of twilight, when limb radiance is formed in dense atmospheric layers below the setting Sun. During the deeper twilight period this difference is reduced, mean $\cos \theta_M$ becomes closer to $\cos \theta_S$. The same can be said about the cosine of mean scattering angle for total NLC field, $\cos \theta_T$.

Particle size analysis is made for solar zenith angles less than 97.5° when NLC are above the shadow of the troposphere. During this twilight stage the difference of $\cos \theta_T$ for the sky point positions 60° and 120° degrees from the Sun is about 0.99975, while this difference for single scattering is equal to unity. Finally, we have a shift of mean particle size $\Delta a/a$ about $10^{-4}$, this is incomparably less than the instrumental error of color measurements.

## 6. Discussion

Multiple scattering is the problem of the sounding of an optically dense atmosphere such as Earth's. Its contribution is sufficient in the twilight background and in the scattered fields of the clouds in the troposphere and stratosphere. The cases when multiple scattering is insignificant are rare, but it is found out to be the highest clouds in the atmosphere, noctilucent clouds.

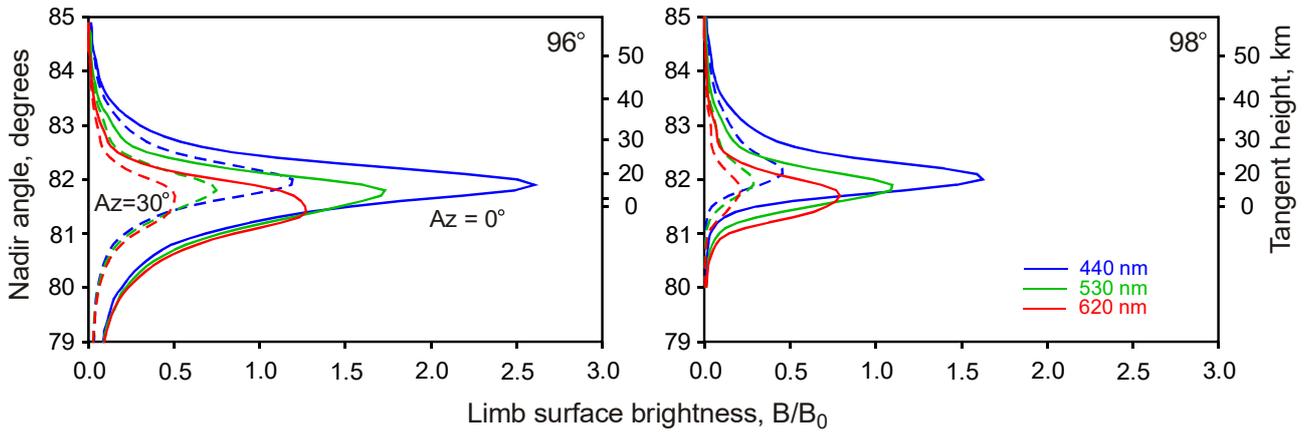

*Figure 11. Radial profiles of the limb radiance at the azimuth angle 0 (near the Sun) and 30°, solar zenith angles 96° and 98°. Tangent heights are calculated with the account of refraction.*

Such a low contribution of multiple scattering in the NLC field during the deep twilight can be checked logically. As we see in Figure 9, the limb of the Earth visible from NLC is a narrow arc. Dependencies of the surface brightness of the limb on the radial coordinate for different solar zenith angles and wavelengths are shown in Figure 11 for the azimuthal angle equal to 0 (direction from the center of the Earth to the Sun) and 30°. Surface brightness $B$ is expressed in the units of $B_0=S_0/4\pi$, where $S_0$ is the unabsorbed brightness of the Sun; $B_0$ is the surface brightness of the solar radiation uniformly scattered in the sky sphere.

The altitude of NLC, 82 km, corresponds to the distance from the Earth's limb about 1000 km. During the period of NLC best visibility (local solar zenith angle 97°) the arc of the solar limb has an effective thickness about 1° or 20 km. It is close to the effective width of the limb for the tangent position of the Sun in LIMBTRAN model (Griffoen and Oikarinen, 2000). The half-maximum length of the arc is about 60°. The effective square of the arc is 60 square degrees or $1.5\cdot10^{-3}$ of the celestial sphere.

It is known that the sky background $B$ cannot be brighter than $B_0$ if the scattering is Lambertian. Since the extinction of the direct solar emission for local solar zenith angle 97° is not strong at the altitude 82 km, the total limb radiance would not be brighter than $(1\text{-}2)\cdot10^{-3}$ of the solar emission in this case. However, the limb radiance in the narrow solid angle close to the Sun position can be brighter by the factor of several units if anisotropic Mie scattering on aerosol particles is dominant, we can see it in Figures 9 and 11. This leads us to the rough estimation of the ratio "limb/Sun" about 1%, the value was fixed by Monte-Carlo analysis for the period of best NLC visibility. The contribution of multiple scattering increases during the light period of twilight $\zeta<96°$ (due to the increase of effective thickness of the arc of the limb radiance), and also during the darker twilight $\zeta>98°$ (due to the extinction of direct solar radiance near the limb).

Limb radiance decreases exponentially with the solar zenith angle, the same does the twilight sky background. Straight solar emission at the altitude of NLC remains almost constant until immersion of the Sun behind the lower atmospheric layers. The stage of maximal brightness ratio "NLC/background" coincides with the maximal contribution of single scattering in the NLC field. This gives the opportunity of the ground-based measurements the altitude and particle size basing on the approximation of single scattering.

## 7. Conclusion

The main purpose of the paper is the estimation of multiple scattering contribution in the observed radiance of the clouds in the middle and upper atmosphere with the focus on the noctilucent clouds. Multiple scattering is found out to be significant in the case of stratospheric clouds, but this fraction almost fades away if the clouds are in the mesosphere and the Sun is above the dense absorbing atmospheric layers.

Ground-based passive altitude and particle size measurements of noctilucent clouds use the data of color indexes of the cloud field. As we see in Figure 8, effects of multiple scattering appear at the local solar zenith angle 98.5°, given the altitude of the clouds is 82 km. This corresponds to the visual border of Earth's shadow on the clouds, where they are hardly seen. Effects lead to overestimation of the "photometric" altitude of NLC by the value not exceeding 0.2 km. This is confirmed by an agreement of photometric and triangulation altitudes of NLC fixed by five-years observations (Ugolnikov et al., 2025). The same can be said about winter noctilucent clouds detected recently at the altitude 70 km (Ugolnikov, 2025).

Errors of the particle size measurements are even smaller, since the technique is based on the NLC analysis far from the Earth's shadow, at the local solar zenith angles about 97°. Multiple scattering is insignificant during this stage, the effective scattering angle on NLC particles is close to the single scattering angle, not bringing a noticeable error to the particle size measurements.

However, multiple scattering must be taken into account in the case of stratospheric clouds and limb satellite measurements of noctilucent clouds, those can be conducted in the wide range of local solar zenith angles, including the daytime. This problem can be solved if space observations are held in Hartley/Huggins UV bands of ozone blocking the scattered background in the lower atmosphere (Robert et al., 2009).